\documentclass[aps,prb,reprint,superscriptaddress,longbibliography,floatfix,showkeys]{revtex4-2}

\usepackage[T1]{fontenc}
\usepackage[utf8]{inputenc}
\usepackage{graphicx}
\usepackage{amsmath,amssymb,amsfonts}
\usepackage{mathtools}
\usepackage{bm}
\usepackage{xcolor}
\usepackage{booktabs}
\usepackage{hyperref}

\begin{document}

\title{Scale-dependent universality class crossover in magnetic skyrmion polymers}

\author{R.~L. Silva}
\email{ricardo.l.silva@ufes.br}
\affiliation{Departamento de Ci\^{e}ncias Naturais, Universidade Federal do Esp\'{i}rito Santo, S\~{a}o Mateus, ES 29932-540, Brazil}

\author{R.~C. Silva}
\email{rodrigo.c.silva@ufes.br}
\affiliation{Departamento de Ci\^{e}ncias Naturais, Universidade Federal do Esp\'{i}rito Santo, S\~{a}o Mateus, ES 29932-540, Brazil}

\author{R.~L. Stamps}
\email[Corresponding author: ]{Robert.Stamps@umanitoba.ca}
\affiliation{Department of Physics and Astronomy, University of Manitoba, Winnipeg, Manitoba R3T 2N2, Canada}

\begin{abstract}
Dipolar magnetic skyrmions can assemble into chains with alternating helicity that act as one-dimensional polymers, yet their statistical mechanics violates the universal harmonic scaling observed in actin, DNA, and microtubules. From first principles, we compute the inter-skyrmion pair potential and find a bi-exponential form of competing interactions with two characteristic decay lengths that encode the distinct microscopic mechanisms of repulsion and attraction. Multiscale simulations reveal a power-law temperature dependence with exponent $1$ in the worm-chain limit of a single bond, and exponent $1/2$ in the three-bond limit. We find that the power-law behavior is remarkably independent of magnetic field strength, and the crossover is due to competing radial interactions responsible for the bonds, resulting in a quartic transverse confinement. We show that the precise form of the competing interactions (e.g., Morse or double-Yukawa) does not affect the temperature dependence.
\end{abstract}


\maketitle


\section{Introduction}\label{sec:intro}

Magnetic skyrmions are topological solitons with particle-like properties that exhibit intriguing physical properties and offer exciting possibilities for applications in spintronics~\cite{Nagaosa2013, Fert2017, Back2020}. The structure and dynamics of these skyrmion chains are, in many respects, analogous to those of chains found in soft matter and are broadly described by similar phenomenological models. On the other hand, there exist unique behaviors of dipolar skyrmion chains that are not observed in other systems. In what follows, we examine a general feature of thermal fluctuations in skyrmion chains that arises from the nature of the interactions stabilizing the chain. We show that this leads to a distinctly different temperature dependence from that observed for soft-matter chains.

\begin{figure*}[t]
    \centering
    \includegraphics[width=\textwidth]{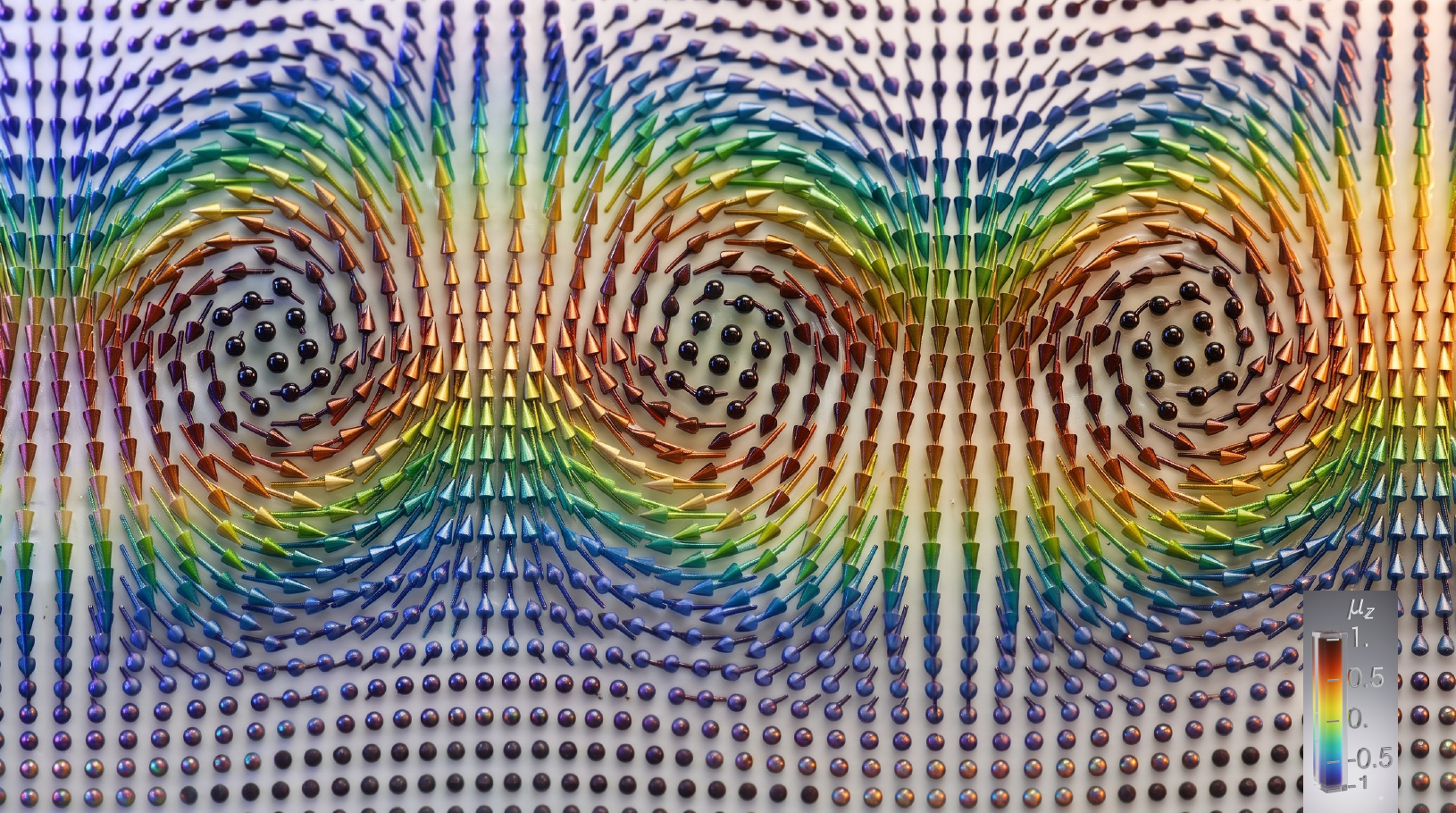}
\caption{\textbf{Skyrmion chain from atomistic simulations.}
    Relaxed spin configuration of a chain of Bloch-type dipolar skyrmions obtained from the atomistic LLG simulations (Methods), embedded in a uniformly magnetized background. Adjacent skyrmions along the chain carry opposite helicities ($\psi = +\pi/2$ and $\psi = -\pi/2$): the in-plane spin texture circulates clockwise around one core and counter-clockwise around the next, and in the junction regions the in-plane moments of adjacent skyrmion walls align in parallel, the real-space signature of the attractive bond between skyrmions of opposite helicity. Magnetic moments are represented by cones colored according to their out-of-plane component $\mu_z$, from $+1$ (up) through $0$ (in-plane) to $-1$ (down), as indicated by the color bar; the in-plane orientation is given by the cone direction.}
    \label{fig:chain}
\end{figure*}

The analogy with soft matter chains is more than qualitative. Skyrmion chains form one-dimensional assemblies bound by a finite-range, helicity-dependent interaction, reminiscent of the case of semi-flexible polymers whose conformations are governed by the balance between bending rigidity and thermal fluctuations. The question addressed here is whether this analogy is reflected by the corresponding balance between interaction and thermal energies in dipolar skyrmion chains. 

In conventional semi-flexible polymers, the bending energy is quadratic in the local curvature, leading to a scaling of angular fluctuations, defined in terms of deflection through an angle $\theta$, which when thermally averaged at temperature $T$, results in $\langle \theta^2 \rangle \propto T$ at all observed length scales. This harmonic scaling is remarkably robust and appears in a wide range of systems. For example, actin filaments are well described by the worm-like chain (WLC) model, whose persistence length is sensitive to the ionic environment and cross-linking. The ionic bonding gives rise to a strictly quadratic bending energy~\cite{Gittes1993}. Furthermore, more complex systems such as microtubules exhibit a strong scale dependence of effective stiffness due to their composite structure, but still preserve the relationship $\langle \theta^2 \rangle \propto T$ at all length scales~\cite{Pampaloni2006}. Similarly, double-stranded DNA, a structurally more complex chain, follows the WLC model for contour lengths greater than the so-called persistence length ($\sim 50$ nm), which describes segment lengths. However, on shorter length scales, strong bending can induce localized kinks that enhance cyclization beyond harmonic predictions~\cite{YanMarko2004, CloutierWidom2004}. Even so, these effects are localized and do not affect global scaling. In regard to other condensed matter systems, we note also that harmonic behavior is observed in dipolar colloidal chains~\cite{deGennes1970, Butter2003, Biswal2003} and vortex lines in type-II superconductors~\cite{Blatter1994}, where quadratic elasticity and linear thermal scaling are preserved.

Across these systems, deviations from harmonicity are absent, as in DNA. In what follows, we show that dipolar skyrmion chains can exhibit fundamentally different scaling as a result of competing interactions that govern their structure. A crossover in thermal scaling from worm-like-chain behavior ($\gamma \approx 1$) on the single-bond scale to quartic statistics ($\gamma \approx 1/2$) on the three-bond scale. This crossover originates from a purely geometric effect: small transverse displacements of a skyrmion produce only a quadratic change in the inter-skyrmion distance, such that a harmonic radial interaction effectively generates a quartic transverse confinement. Remarkably, the effect depends only on the local curvature of the interaction potential at equilibrium and is independent of its global shape. To the best of our knowledge, no molecular or colloidal polymer exhibits such a crossover in the exponent governing transverse thermal fluctuations, establishing skyrmion chains as a unique platform for mechanical spectroscopy of topological solitons.


\section{Results}\label{sec:results}

The unique and complex spin textures in skyrmion chains result from a competition between short-range exchange interactions and local crystal field effects that determine preferred orientation directions for local spin moments. To understand skyrmion structure, it is helpful to first introduce how chirality and size are defined: skyrmions are characterized by a chirality, which defines the direction that spin moments align around the skyrmion core, and by the radial size of the structure. For chiral skyrmions, chirality is governed by a so-called Dzyaloshinskii-Moriya interaction (DMI). This interaction locks the helicity to the crystal symmetry and renders the pair interaction isotropic and purely repulsive~\cite{Ross2021, Capic2020, Lin2013, Brearton2021}. In contrast, the structure of dipolar skyrmions is instead defined by magnetic flux closure of long-range magnetostatic fields~\cite{MoreauLuchaire2016, Woo2016}.
Dipolar skyrmions are stabilized by the competition between the perpendicular uniaxial anisotropy, which favors uniform out-of-plane magnetization, and the magnetostatic (shape) energy, which favors flux closure. This balance is quantified by the quality factor $Q_f = K_u/K_d$, with $K_d = \mu_0 M_s^2/2$ the shape anisotropy; dipolar skyrmions are stable for $Q_f \approx 1$, and the [Co/Ni]$_5$ parameters used here give $Q_f \approx 1.04$, inside the stability window identified experimentally in the same material system~\cite{Hassan2024}. The helicity of a dipolar skyrmion is not fixed by any microscopic interaction: Bloch textures of either rotation sense are degenerate in energy, so the helicity is a soft internal degree of freedom. Importantly, right- and left-handed skyrmions can exist in these materials without preference because there is no intrinsic dependence of chirality on crystalline symmetries. As a consequence, the energy of pair interactions between skyrmions can be minimized by alternating helicity between neighboring skyrmions: the in-plane circulations of adjacent skyrmion walls then mesh smoothly in the junction regions, minimizing the exchange cost of wall overlap (Fig.~\ref{fig:chain}). For this reason, dipolar skyrmions can form stable, cohesive linear chains~\cite{Du2015, Hassan2024, Jefremovas2025}.

\subsection{Micromagnetic origin of the inter-skyrmion interaction}\label{sec:micro}

The pair interaction between dipolar skyrmions has a rigorous micromagnetic foundation that directly motivates a bi-exponential form we propose as a model. The model consists of two parts. The short-ranged exchange and local anisotropy energies define the individual skyrmion structure, which we assume is relatively rigid, given the large energies associated with deformation. Dipolar magnetostatic energies are, in comparison, significantly weaker but are long-range. 


The essential idea is that two channels of different physical origin compete. The repulsion is magnetostatic: the in-plane texture of a Bloch skyrmion is purely tangential and therefore divergence-free, so it carries no volume magnetic charges, and the magnetostatic interaction of rigid profiles reduces to the helicity-independent surface-charge channel, which is repulsive at all separations; at short range it is supplemented by the overlap repulsion of the deformed domain-wall profiles. The attraction is exchange-mediated and helicity-selective: the cross exchange energy of the overlapping wall textures carries the product of the two chiralities and is attractive precisely for skyrmions of opposite helicity, whose in-plane circulations mesh smoothly in the junction region (Fig.~\ref{fig:chain}). The equilibrium bond length ($r_\mathrm{eq}$) is set by the balance between this junction-mediated attraction and the repulsive channels. The isolated-skyrmion structure, in turn, is characterized by its radius ($R_0$) and domain-wall width ($\delta_w$). In what follows, we derive both channels from a Fourier--Bessel treatment and use these distinct physical mechanisms to motivate the effective pair potential tested against numerical micromagnetic simulations.

A skyrmion with helicity $\gamma$ and topological charge $Q = -1$ has magnetization
\begin{equation}
\mathbf{m}(\rho) = \left(\sin\Theta\cos(\phi+\gamma),\;\sin\Theta\sin(\phi+\gamma),\;\cos\Theta\right),
\label{eq:magnetisation}
\end{equation}
where $\Theta(\rho)$ is the polar angle at radial distance $\rho$ from the center of the skyrmion and $\phi$ is the azimuthal angle. The dipolar skyrmions of interest here are Bloch textures, $\gamma = \pm\pi/2$, the helicity selected by the magnetostatic self-energy in the absence of DMI (Supplementary Note~S4): the radial component vanishes, $m_\rho = \cos\gamma\sin\Theta = 0$, and the in-plane magnetization is a purely tangential circulation, $\mathbf{m}_\parallel = c\,\sin\Theta(\rho)\,\hat{\bm{\phi}}$, with $c \equiv \sin\gamma = \pm 1$ the chirality. We approximate the radial spin configuration with a domain-wall ansatz $\Theta(\rho) = 2\arctan\!\left[e^{-(\rho - R_0)/\delta_w}\right]$, which gives $\sin\Theta = \mathrm{sech}[(\rho - R_0)/\delta_w]$ and $\cos\Theta = \tanh[(\rho - R_0)/\delta_w]$, where the equilibrium radius $R_0$ is assumed to be determined by the parameter $\delta_w$ which in one dimensional magnetic solitons corresponds to a domain-wall width.

For the magnetostatic energy, we need to consider two types of magnetic charge: surface and volume.
Surface charges are defined by $\sigma_\mathrm{top} = +M_s m_z$ where $M_s$ is the saturation magnetization. The surface charge depends only on the $m_z$ component of the magnetization and is independent of the helicity and of the chirality.
Volume charges, $\rho_\mathrm{vol} = -M_s\,\nabla \cdot \mathbf{m}_\parallel$, carry a factor $\cos\gamma$ and \emph{vanish identically} for the Bloch texture: the purely tangential in-plane magnetization is divergence-free, for any radial profile (Supplementary Note~S2). The magnetostatic interaction between Bloch skyrmions is therefore carried entirely by the surface charges, and is blind to the chiralities --- it cannot select the helicity order of the chains.

The 2D Fourier transforms of the two magnetization components are obtained via a Fourier-Bessel (Hankel) analysis as described in the supplemental material and described by a Hankel function $G_n(\mathbf{k})$ where $\mathbf{k}$ is the $k$-space transform vector and $n$ is the order.
The out-of-plane part is written as $\delta m_z \equiv m_z - 1$ for the excess charge relative to the uniform background, which yields
\begin{equation}
\widetilde{\delta m}_z(\mathbf{k}) = 2\pi\, G_0(\mathbf{k}).
\label{eq:FT_mz}
\end{equation}
This transform is isotropic and helicity-independent.
The in-plane magnetization decomposes into a longitudinal (charge-carrying) projection along $\hat{k}$ and a transverse (divergence-free) projection along $\hat{k}_\perp = \hat{z}\times\hat{k}$:
\begin{align}
\hat{k}\cdot\tilde{\mathbf{m}}_\parallel(\mathbf{k})
    &= -2\pi i\,\cos\gamma\;G_1(k) = 0, \nonumber\\
\tilde{\mathbf{m}}_\parallel(\mathbf{k})
    &= -2\pi i\,c\;G_1(k)\,\hat{k}_\perp .
\label{eq:FT_inplane}
\end{align}
where the longitudinal projection vanishes at the Bloch helicity --- the reciprocal-space form of the vanishing of the volume charges. The surviving transverse projection carries no magnetic charge and generates no stray field: it is magnetostatically silent, but remains fully visible to the exchange energy, which is where the helicity selection ultimately resides.

For a thin film of thickness $D$, the surface-charge demagnetizing energy carries the exact finite-thickness kernel $h_s(kD) = \left(1 - e^{-kD}\right)/(kD)$, with $h_s \approx 1$ in the thin-film limit (Supplementary Note~S4).
For two Bloch skyrmions separated by a distance $R$ in the film plane, the magnetostatic interaction energy is then purely the surface-charge term,
\begin{equation}
\begin{split}
E_s(R) ={}& 2\pi\mu_0 M_s^2 D \int_0^\infty dk\,k\;
 h_s(kD)\\
&\times G_0^2(k)\,J_0(kR) > 0 .
\end{split}
\label{eq:Eint}
\end{equation}
which is repulsive at all separations and exactly degenerate in the chiralities $(c_1, c_2)$: it supplies the repulsive channel of the pair potential, but neither the binding nor the helicity order can be magnetostatic in origin.
The attraction is carried by the remaining leading interaction of the overlapping textures: the cross exchange energy. Within the superposition of two Bloch skyrmions it factorizes into the same two Hankel transforms (Supplementary Note~S8),
\begin{equation}
\begin{split}
E^\mathrm{ex}_\mathrm{int}(R) ={}& 4\pi A_\mathrm{ex} D
\int_0^\infty dk\,k^3\\
&\times \left[G_0^2(k)+c_1c_2G_1^2(k)\right]J_0(kR) .
\end{split}
\label{eq:Eex}
\end{equation}
and carries the chirality product $c_1 c_2$ --- the only term among the leading interactions that does. The in-plane contribution is attractive precisely for opposite helicities ($c_1 c_2 = -1$): on the segment between the two centers the azimuthal unit vectors of the two skyrmions are antiparallel, so opposite chiralities make the in-plane circulations locally parallel in the junction --- two counter-rotating gears meshing smoothly --- lowering the exchange cost of the overlap, whereas equal chiralities force the circulations into a head-on collision. This is the microscopic mechanism for helicity-alternation binding in dipolar skyrmion chains, and its real-space signature is directly visible in the relaxed chains: the in-plane moments of adjacent walls align in parallel across the junction regions (Fig.~\ref{fig:chain}).

The decay behavior of both channels follows from the analytic structure of the transforms in the complex $\rho$-plane, where the domain-wall profile places a ladder of poles at $\rho_n = R_0 + i(n+\tfrac{1}{2})\pi\delta_w$. Supplementary Note~S6 carries out the full contour integration for the surface channel: the $\mathrm{sech}^2$ wall derivative produces \emph{double} poles whose equal-sign residues resum into a hyperbolic-cosecant envelope,

\begin{equation}
G_0(k) \simeq -\pi\delta_w\sqrt{\tfrac{2R_0}{\pi k}}\, \cos\!\left(kR_0 - \tfrac{3\pi}{4}\right) \mathrm{csch}\!\left(\tfrac{\pi k\delta_w}{2}\right),
\label{eq:G0-closed}
\end{equation}
\noindent
so that the wall width $\delta_w$ --- the distance of the dominant pole from the real axis --- is the intrinsic length scale of the interaction. The magnetostatic repulsion~(\ref{eq:Eint}) then decays with a strict dipolar tail $+\mu_0\mu_\mathrm{sk}^2/(4\pi R^3)$ at $R \gg R_0$ (the two skyrmion cores repel as parallel point dipoles), but over the physically relevant range $R_0 < R < 4R_0$, where the numerical micromagnetic pair potential is measured, it is accurately described by a single effective exponential of range $\lambda_s \simeq 1.2\,\delta_w$ (Supplementary Note~S7):
\begin{equation}
E_s(R) \approx +B_s\,e^{-R/\lambda_s}.
\label{eq:repulsion}
\end{equation}
The attractive channel~(\ref{eq:Eex}) is longer ranged. Its asymptotic decay is $e^{-(R-2R_0)/\delta_w}$, set by the overlap of the in-plane wall tails, but it carries a polynomially growing junction-area prefactor: about half of the attraction accumulates in the junction between the skyrmions, and the local effective decay length drifts across $(2$--$4)\,\delta_w$ in the window sampled by bound neighbors (Supplementary Note~S8),
\begin{equation}
\begin{split}
E^\mathrm{ex}_\mathrm{int}(R) &\approx -C_\mathrm{att}\,e^{-R/\lambda},\\
\lambda &\sim (2\text{--}4)\,\delta_w
\qquad (c_1c_2=-1).
\end{split}
\label{eq:attraction}
\end{equation}

\subsection{Bi-exponential pair potential}\label{sec:biexp}

The total pair potential $U_{total}(r)$ combines the junction-mediated exchange attraction derived above with the surface-charge magnetostatic repulsion of Eq.~(\ref{eq:repulsion}) and, at short range, the Slonczewski domain-wall overlap repulsion\cite{Malozemoff1979} arising from the overlap of the exponential tails of the two walls, which shares the same wall-width scale. The dominant terms are two exponentials: a repulsive one that decays as $\delta \sim \delta_w$ and an attractive one that decays as $\lambda \sim (2\text{--}4)\,\delta_w$. We can thus describe the dominant contributions to the total potential as an effective potential $U(r)$ postulated as a sum over two competing exponentials:
\begin{equation}
U(r) = B_\mathrm{rep}\,e^{-r/\delta} - C_\mathrm{att}\,e^{-r/\lambda}, \quad \lambda > \delta.
\label{eq:biexp}
\end{equation}
The coefficients $B_\mathrm{rep}$ and $C_\mathrm{att}$ are amplitudes describing the relative strengths of the two terms, and the decay rates describe the relative lengthscales involved. The hierarchy $\lambda > \delta$ reflects the distinct spatial structure of the two channels: the repulsion is fed by the localized wall derivative and the fast $\delta_w/2$ tail of $\delta m_z$, while the attraction accumulates over the extended junction region between the two walls, which stretches its effective range.
Supplementary Note~S9 shows that this inequality is a direct consequence of the distinct physical origins of the two channels rather than a free assumption, and that the attractive exponential exists only in the opposite-helicity channel $c_1 c_2 = -1$ --- the bi-exponential potential and the alternating-helicity order of the chains follow from one and the same selection rule. The well-known Morse potential corresponds to the special case $\lambda = 2\delta$.

To extract the parameters in Eq.~(\ref{eq:biexp}), we perform atomistic Landau-Lifshitz-Gilbert (LLG) simulations of [Co/Ni]$_5$ multilayers (see Methods).
Two skyrmions with opposite helicities are initialized at different center-to-center distances $r$, and then relaxed towards equilibrium by over-damped LLG dynamics. The effective pair potential is obtained by subtracting the self-energy of two isolated skyrmions\cite{Ross2021,Capic2020} such that $U(r) = E_\mathrm{tot}(r) - 2E_\mathrm{sky}$.

Table~\ref{tab:biexp} shows the fitted parameters for three magnetic field strengths.
The decay-length ratio $\lambda/\delta \approx 1.5$ is remarkably independent of the field strength, consistent with the micromagnetic prediction that both decay lengths are set by the same microscopic scale --- the domain-wall width, which the field changes only weakly --- rather than by the skyrmion radius, which it changes substantially: $\delta$ is governed by the surface-charge repulsion and wall overlap, and $\lambda$ by the junction-mediated exchange attraction (Supplementary Note~S9). At $B = 35$~mT the fitted values, $\delta = 10.5$~nm~$= 1.6\,\delta_w$ and $\lambda = 15.8$~nm~$= 2.4\,\delta_w$, fall inside the analytically predicted ranges.
The fits yield a coefficient of determination $R^2$ ranging from $0.97$ to $0.98$, indicating a faithful representation of the numerical micromagnetic model.

\begin{table*}[t]
\centering
\caption{\textbf{Bi-exponential fit parameters from LLG simulations.} The decay-length ratio $\lambda/\delta \approx 1.5$ is stable across all fields, reflecting that both decay lengths are set by the domain-wall width: $\delta$ by the surface-charge repulsion and wall overlap, $\lambda$ by the junction-mediated exchange attraction. The Morse potential corresponds to the special case $\lambda = 2\delta$.}
\label{tab:biexp}
\begin{tabular}{lccc}
\toprule
 & $B = 30$~mT & $B = 35$~mT & $B = 40$~mT \\
\midrule
$B_\mathrm{rep}$ ($J_1$) & 74.20 & 63.36 & 42.75 \\
$C_\mathrm{att}$ ($J_1$) & 48.63 & 43.72 & 33.07 \\
$\delta$ (nm) & 11.68 & 10.54 & 9.89 \\
$\lambda$ (nm) & 17.53 & 15.80 & 14.84 \\
$r_\mathrm{eq}$ (nm) & 29.1 & 24.5 & 19.7 \\
$\lambda/\delta$ & 1.50 & 1.50 & 1.50 \\
$D_e$ ($J_1$) & 3.10 & 3.07 & 2.94 \\
$U''(r_\mathrm{eq})$ ($J_1$/nm$^2$) & 0.0149 & 0.0187 & 0.0198 \\
$R^2$ & 0.97 & 0.98 & 0.97 \\
\bottomrule
\end{tabular}
\end{table*}

\subsection{Transverse confinement and fluctuations}\label{sec:quartic}

\begin{figure}[t]
    \centering
    \includegraphics[width=0.95\columnwidth]{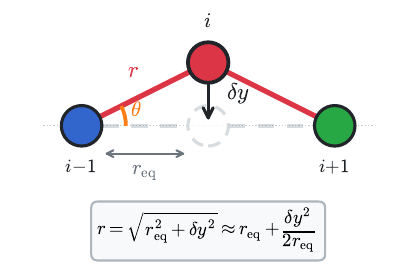}
    \caption{\textbf{Geometric coupling mechanism.}
    A transverse deflection $\delta y$ stretches both bonds from $r_\mathrm{eq}$ to $r = \sqrt{r_\mathrm{eq}^2 + \delta y^2}$.
    The resulting energy cost scales as $(\delta y)^4$, not $(\delta y)^2$.}
    \label{fig:geometry}
\end{figure}

We now discuss the central result of this work: fluctuations in the $x-y$ plane around equilibrium spacing for the skyrmion chain. For an equilibrium chain aligned along the $x$ axis, the thermal fluctuations of the chain correspond to the transverse displacements $\delta y$ of individual skyrmions perpendicular to the chain axis. These correspond to a change in the radius of the skyrmion $r$ away from $R_0$. From  Eq.~(\ref{eq:biexp}) it is clear that $-\frac{\partial U}{\partial r}$ represents a purely radial central force between point-like objects.  
Such displacements deform the skyrmion chain and can be thought of as stretching bonds between neighboring skyrmions. Note that these radial deformations couple transverse and longitudinal degrees of freedom, as can be inferred from Fig.~\ref{fig:geometry}.   

Consider three consecutive skyrmions, the central one displaced transversely by $\delta y$ while its neighbors remain fixed (Fig.~\ref{fig:geometry}).
The bond length becomes
\begin{equation}
r = \sqrt{r_\mathrm{eq}^2 + \delta y^2} \approx r_\mathrm{eq} + \frac{\delta y^2}{2r_\mathrm{eq}} - \frac{\delta y^4}{8r_\mathrm{eq}^3} + \mathcal{O}(\delta y^6).
\label{eq:bond_stretch}
\end{equation}
Crucially, the stretch of the bond $\Delta r \equiv r - r_\mathrm{eq}$ starts on the order $\delta y^2$.
The expansion of the central potential $U(r)$ to its minimum gives
\begin{equation}
U = U(r_\mathrm{eq}) + \tfrac{1}{2}U''(r_\mathrm{eq})\,\Delta r^2 + \tfrac{1}{6}U'''(r_\mathrm{eq})\,\Delta r^3 + \cdots.
\end{equation}
Noting that $(\Delta r)^n$ starts at $\delta y^{2n}$, we find that only the $n = 2$ term contributes at order $\delta y^4$: the cubic radial term enters at $\delta y^6$ and the quartic radial term at $\delta y^8$.
The correction $\delta y^4$ in Eq.~(\ref{eq:bond_stretch}) itself produces only $\delta y^6$ terms through the cross product in $\Delta r^2$.
The single-bond result is therefore exactly to order $\delta y^4$:
\begin{equation}
U_\mathrm{geom}(\delta y) = U(r_\mathrm{eq}) + \frac{U''(r_\mathrm{eq})}{8\,r_\mathrm{eq}^2}\,(\delta y)^4 + \mathcal{O}(\delta y^6).
\label{eq:quartic}
\end{equation}
In this way, the lowest order correction to the skyrmion chain radial potential is order four in the transverse direction. This is a crucial distinction from classical WLC polymers, where the lowest order bending is quadratic in transverse displacement. Note also that in general the relative displacements of the three skyrmions shown in Fig.~\ref{fig:geometry} will result in the pairwise separations being unequal. However, for small $\delta y$, the leading term transverse corrections to the energy remain the same as in Eq.~(\ref{eq:quartic}) with displacements along the $x$ direction entering as longitudinal fluctuations to quadratic order.

Adding the explicit WLC bending term extracted from the LLG simulations (see Methods), the effective local potential becomes
\begin{equation}
U_\mathrm{eff}(\delta y) = \underbrace{\frac{U''(r_\mathrm{eq})}{4r_\mathrm{eq}^2}(\delta y)^4}_{\text{quartic (geometric)}} + \underbrace{\frac{\kappa_\mathrm{WLC}}{2r_\mathrm{eq}^2}(\delta y)^2}_{\text{harmonic (WLC)}}.
\label{eq:full_potential}
\end{equation}
This effective potential has the form of a Landau $\varphi^4$ potential with $\delta y$ playing the role of the order parameter.
The analogy is structural rather than thermodynamic: here $\kappa_\mathrm{WLC}$ is always positive and no phase transition occurs, but the crossover from quadratic to quartic dominance is governed by the same change in the degree of the confining potential that, in equilibrium statistical mechanics, dictates different thermal exponents ($\gamma = 1/n$ for $U \propto |x|^{2n}$).
For a purely quartic potential $U = \lambda x^4$, dimensional analysis gives $\langle x^2 \rangle \propto (k_BT/\lambda)^{1/2} \propto T^{1/2}$, in contrast to the harmonic result $\langle x^2 \rangle \propto T$ from equipartition.

Using $U''(r_{\rm eq}) = 0.0187~J_1/\mathrm{nm}^2$ from the bi-exponential
fit (Table~\ref{tab:biexp}) together with the bending rigidity
$\kappa_{\rm WLC} = 1.19~J_1$ extracted from the static arc-bending
protocol (see Methods), we obtain the following results.
\begin{equation}
\begin{split}
\delta y_{\rm cross}
&= \sqrt{\frac{2\kappa_{\rm WLC}}{U''(r_{\rm eq})}}
 \approx 11~\text{nm},\\
&\hspace{1.2cm} B=35~\text{mT},\quad T=300~\text{K}.
\end{split}
\label{eq:dycross}
\end{equation}
When $\sqrt{\langle \delta y^2 \rangle} \ll \delta y_\mathrm{cross}$, the harmonic term dominates and $\gamma \approx 1$; when $\sqrt{\langle \delta y^2 \rangle} \gg \delta y_\mathrm{cross}$, the quartic term prevails and $\gamma \approx 1/2$.

\subsection{Scale-dependent thermal exponents}\label{sec:thiele}
\begin{figure*}[t]
    \centering
    \includegraphics[width=\textwidth]{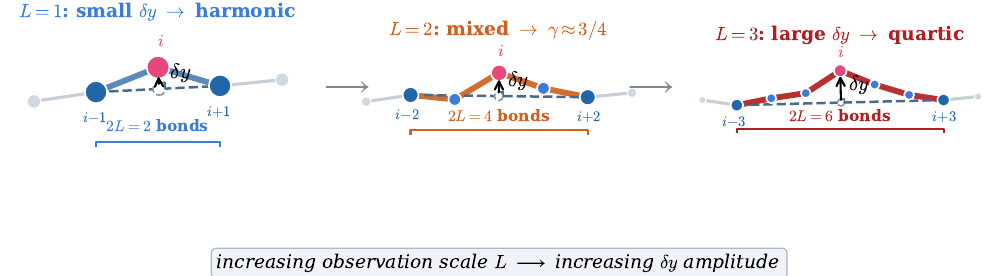}
    \caption{Scale-dependent crossover mechanism.
    At $L = 1$, harmonic bending dominates ($\gamma \approx 1$).
    At $L = 2$, both terms contribute ($\gamma \approx 3/4$).
    At $L = 3$, quartic geometric confinement dominates ($\gamma \approx 1/2$).}
    \label{fig:geometry_scales}
\end{figure*}

When quartic order contributions to the transverse fluctuations dominate, the temperature-scaling exponent $\gamma$ in $\langle \delta y^2(L) \rangle \propto T^\gamma$ should depend on the chain lengths over which observational averages are made. This idea is illustrated in Fig.~\ref{fig:geometry_scales}. We define $L$ as the chain length, or window, over which transverse fluctuation measurements are made. From the figure, one sees that the window $L$ corresponds to the number of skyrmion bonds observed. Larger observation windows accumulate larger transverse excursions, and thus we expect a scaling of fluctuation magnitude with window size. 

Because thermally driven dynamics for atomistic micromagnetics is computationally demanding for very large chains, simulations of a stochastic Thiele model were performed instead in order to obtain ensemble averages for transverse displacements over long skyrmion chains.\cite{Thiele1973} Our implementation enabled simulations of chains containing $N = 200$ skyrmions and was based on numerical integration of
\begin{equation}
	\mathbf{G} \times \dot{\mathbf{R}}_i + \alpha_D \dot{\mathbf{R}}_i = -\nabla_{\mathbf{R}_i} E_\mathrm{tot} + \bm{\eta}(t),
	\label{eq:thiele}
\end{equation}
where $\mathbf{G} = G_z \hat{z}$ is the gyrovector ($G_z = 4\pi$) and $\alpha_D$ the dissipation coefficient.
The thermal noise $\bm{\eta}(t)$ satisfies $\langle \eta_i^\mu(t) \eta_j^\nu(t') \rangle = 2\alpha_D k_B T \delta_{ij}\delta_{\mu\nu}\delta(t-t')$\cite{Miltat2018,Schutte2014}.
All magnetic parameters are kept fixed at their room-temperature values; temperature enters exclusively through the stochastic force amplitude.
We sample 10 temperatures from 240 to 510~K with 256 independent samples per temperature, yielding 7680 total samples across all three field strengths (see Methods for the full simulation protocol).
Once the bi-exponential parameters, $\kappa_\mathrm{WLC}$, and $\alpha_D$ are extracted from the LLG calculations, the Thiele simulations do not introduce additional temperature-dependent fitting parameters.

\begin{figure*}[t]
    \centering
    \includegraphics[width=\textwidth]{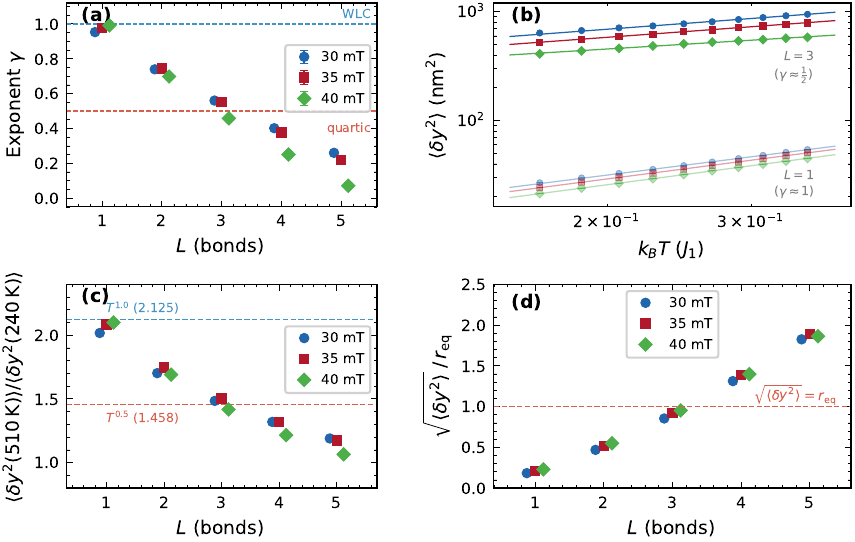}
    \caption{\textbf{Scale-dependent thermal-scaling crossover.}
    \textbf{a,} Temperature scaling exponent $\gamma$ versus segment length $L$ at $B = 30$~mT (circles), 35~mT (squares), and 40~mT (diamonds).
    Dashed lines mark $\gamma = 1$ (WLC) and $\gamma = 1/2$ (quartic).
    \textbf{b,} $\langle \delta y^2 \rangle$ versus $k_B T$ for $L = 1$ and $L = 3$ at all three fields, showing the distinct slopes (exponents) at the two scales.
    \textbf{c,} Temperature-ratio test: $\langle \delta y^2(510\,\mathrm{K})\rangle / \langle \delta y^2(240\,\mathrm{K})\rangle$ versus $L$.
    All three fields show $L = 1$ tracking the $T^{1.0}$ prediction and $L = 3$ approaching $T^{0.5}$; the drop below $1/2$ at $L \geq 4$ signals potential saturation.
    \textbf{d,} Universal geometric criterion: $\sqrt{\langle \delta y^2 \rangle}/r_\mathrm{eq}$ at $T = 300$~K versus $L$.
    The crossover to quartic statistics ($\gamma \approx 1/2$) occurs at $L = 3$, where $\sqrt{\langle \delta y^2 \rangle} \approx r_\mathrm{eq}$ (dashed line), independently of the magnetic field strength.
    Data: $N = 200$, $G_z = 4\pi$, 7680 samples total.}
    \label{fig:crossover}
\end{figure*}

The results are shown in $B = 35~\text{mT}$ and $T = 300~\text{K}$ in Fig.~\ref{fig:crossover}d. At $L = 1$, the symmetric window spans two bonds ($i{-}1$ to $i{+}1$) and the transverse displacement of the central skyrmion is related to the angle of the bond by $\delta y \approx r_\mathrm{eq}\,\theta/2$; the measured amplitude $\sqrt{\langle \delta y^2 \rangle} \approx 6$~nm $\ll \delta y_\mathrm{cross}$ confirms the harmonic regime and one expects $\gamma \approx 1$.
At $L = 2$, the window spans four bonds ($i{-}2$ to $i{+}2$) and the central skyrmion's RMS deflection reaches $\sqrt{\langle \delta y^2 \rangle} \approx 14$~nm, comparable to $\delta y_\mathrm{cross}$, placing the chain in the mixed regime with $\gamma \approx 3/4$.
At $L = 3$, the window spans six bonds ($i{-}3$ to $i{+}3$) and the central skyrmion reaches $\sqrt{\langle \delta y^2 \rangle} \approx 24$~nm $\gg \delta y_\mathrm{cross}$, so the quartic term dominates and $\gamma \approx 1/2$.
Thus, the progression $L=1 \rightarrow 2 \rightarrow 3$ is directly mapped onto the harmonic confinement $\rightarrow$ mixed $\rightarrow$ quartic (Fig.\ref{fig:geometry_scales}).
Operationally, the crossover coincides with $\sqrt{\langle \delta y^2 \rangle} \approx r_\mathrm{eq}$: the quartic regime emerges once transverse fluctuations become comparable to the bond length itself(Fig.\ref{fig:crossover} d).
This criterion is verified across all three field strengths in the simulations that follow, where the ratios $\sqrt{\langle \delta y^2(L{=}3) \rangle}/r_\mathrm{eq}$ at 300~K cluster near unity (0.92, 0.99, and 1.09 for $B = 30$, 35 and 40~mT, respectively), providing a simple geometric rule: $\gamma \approx 1/2$ when $\sqrt{\langle \delta y^2 \rangle} \gtrsim r_\mathrm{eq}$.
Table~\ref{tab:comparison} summarizes the two limiting regimes.

\begin{table}[hbt]
\caption{Scaling comparison between the two universality classes present in skyrmion chains.}
\label{tab:comparison}
\begin{tabular}{lcc}
Property & WLC ($L = 1$) & Quartic ($L \geq 3$) \\
\hline
Effective potential & $\propto \theta^2$ & $\propto (\delta y)^4$ \\
$\langle \delta y^2 \rangle$ & $\propto T^{1.0}$ & $\propto T^{0.5}$ \\
Physical origin & Bending rigidity & Geometric coupling \\
\end{tabular}
\end{table}

The mean-square transverse fluctuation $\langle \delta y^2(L,T) \rangle$ is evaluated for observation scales $L = 1, 2, 3, 4,$ and $5$ by averaging on all internal sites and all samples at each temperature. Operationally, for each skyrmion~$i$ we construct a symmetric window of half-width $L$ that spans skyrmions $i{-}L$ to $i{+}L$ ($2L$ bonds), draw the chord connecting the two endpoints, and define $\delta y$ as the perpendicular distance of the central skyrmion~$i$ from that chord(See Methods).
Log-log fits of $\langle \delta y^2 \rangle = A\,T^\gamma$ yield the exponent $\gamma(L)$ shown in Fig.~\ref{fig:crossover}a and ~\ref{fig:crossover}b.
At $B = 30$~mT, $\gamma$ decreases from $0.953 \pm 0.008$ at $L = 1$ to $0.739 \pm 0.011$ at $L = 2$ and $0.560 \pm 0.012$ at $L = 3$.
The same progression appears at $B = 35$~mT, with $\gamma = 0.977 \pm 0.005$ at $L = 1$, $0.745 \pm 0.004$ at $L = 2$, and $0.551 \pm 0.004$ at $L = 3$.
As a third independent validation, $B = 40$~mT gives $\gamma = 0.991 \pm 0.003$ at $L = 1$, $0.698 \pm 0.008$ at $L = 2$, and $0.458 \pm 0.012$ at $L = 3$.

The intermediate result $L = 2$ is neither a small correction to the WLC behavior nor an already asymptotic quartic scaling, but precisely the mixed-confinement regime expected from Eq.~(\ref{eq:full_potential}).
The measured $\gamma(L{=}2)$ values, 0.739, 0.745, 0.698 for $B = 30$, 35, and 40~mT, cluster in the range 0.70--0.75, within 1--7\% of the arithmetic mean $(1 + 1/2)/2 = 3/4$.
Beyond the quartic regime, $\gamma$ continues to decrease: at $L = 4$, $\gamma \approx 0.25$--$0.40$, already below $1/2$, signaling the onset of potential saturation as the system explores 50--60\% of the dissociation energy.

A complementary, model-independent study is provided by a temperature-ratio test shown in Fig.~\ref{fig:crossover}c.
For $T_\mathrm{high}/T_\mathrm{low} = 510/240 = 2.125$, the predicted fluctuation ratio is $2.125^{1.0} = 2.125$ for the behavior of WLC and $2.125^{0.5} = 1.458$ for the quartic statistics.
At $B = 35$~mT, the measured ratios are 2.084 at $L = 1$ and 1.505 at $L = 3$; at $B = 30$~mT, they are 2.018 and 1.485, respectively; and at $B = 40$~mT, they are 2.100 and 1.417, respectively.
Because this comparison uses raw fluctuation ratios rather than a free exponent fit, it directly confirms that the two limiting responses are already present in the data.

The crossover also survives restriction to the experimentally realized temperature window (240--300~K).
Restricting the power-law fits to only three temperatures within this window still yields clearly separated exponents: at $B = 40$~mT, $\gamma = 0.973 \pm 0.001$ at $L = 1$ and $\gamma = 0.504 \pm 0.007$ at $L = 3$.
The temperature-ratio test is equally robust: $\langle \delta y^2(300\,\text{K}) \rangle / \langle \delta y^2(240\,\text{K}) \rangle$ gives 1.242 in $L = 1$ (prediction for $\gamma = 1$: 1.250) and 1.120 in $L = 3$ (prediction for $\gamma = 1/2$: 1.118), confirming the dual scaling within the experimentally accessible window.


The quartic transverse confinement arises from the geometric identity $\Delta r \approx \delta y^2/(2r_\mathrm{eq})$ applied to the harmonic radial term $\frac{1}{2}U''(r_\mathrm{eq})\,\Delta r^2$.
Because only $U''(r_\mathrm{eq})$ enters in the leading order, the crossover is independent of the global shape of the pair potential.
The crossover can fail only if: (i) $U''(r_\mathrm{eq}) = 0$ (degenerate minimum); (ii) the potential is explicitly angular-dependent; or (iii) the interacting units have a composite internal structure.

We verified this behavior numerically by replacing the bi-exponential potential with both a Morse potential and a double-Yukawa potential fitted to the same LLG data.
Table~\ref{tab:potentials} compares the thermal scaling exponents at $B = 30$~mT.
All three potentials produce the same crossover sequence with $|\Delta\gamma| < 0.03$ and no difference exceeding $1.5\sigma$ on any scale.

\begin{table*}[t]
\centering
\caption{\textbf{Potential-form independence of the thermal scaling exponent.} Values of $\gamma$ from Thiele simulations at $B = 30$~mT using bi-exponential, Morse, and double-Yukawa potentials fitted to the same LLG data.}
\label{tab:potentials}
\begin{tabular}{lccc}
\toprule
$L$ & $\gamma_\mathrm{bi\text{-}exp}$ & $\gamma_\mathrm{Morse}$ & $\gamma_\mathrm{Yukawa}$ \\
\midrule
1  & $0.953 \pm 0.008$ & $0.888 \pm 0.005$ & $0.899 \pm 0.006$ \\
2  & $0.739 \pm 0.011$ & $0.651 \pm 0.010$ & $0.672 \pm 0.012$ \\
3  & $0.560 \pm 0.012$ & $0.470 \pm 0.013$ & $0.495 \pm 0.016$ \\
5  & $0.260 \pm 0.012$ & $0.199 \pm 0.016$ & $0.223 \pm 0.019$ \\
10 & $-0.222 \pm 0.008$ & $-0.190 \pm 0.014$ & $-0.190 \pm 0.014$ \\
\bottomrule
\end{tabular}
\end{table*}



\section{Discussion}\label{sec:discussion}

The crossover $\gamma(L)$ is a local property encoded in the interaction potential and therefore does not require a long chain to exist.
What requires larger systems ($N \gtrsim 50$) is the measurement of the crossover, since one needs enough internal sites to average $\langle \delta y^2(L) \rangle$ accurately.
Current imaging experiments report dipolar skyrmion chains with up to $N \sim 10$\cite{Hassan2024,Jefremovas2025}; longer chains should be accessible through field control, substrate patterning, or material optimization.

Although the existence of the $\gamma(L)$ crossover is universal, its location encodes the interaction.
The crossover scale $\delta y_\mathrm{cross} = \sqrt{2\kappa_\mathrm{WLC}/U''(r_\mathrm{eq})}$ and the quartic coefficient $\lambda_q = U''(r_\mathrm{eq})/(4r_\mathrm{eq}^2)$ depend only on the curvature of the pair potential at equilibrium, not on its global shape.
Measurement $\gamma(L)$ from thermal skyrmion-position maps thus amounts to a mechanical spectroscopy of the pair interaction, complementary to direct energy measurements from static force--distance curves.

A particularly direct experimental test follows from the dual scaling.
The ratio $\langle \delta y^2(L{=}3) \rangle / \langle \delta y^2(L{=}1) \rangle$ should decrease as $T^{-1/2}$ with increasing temperature, since the numerator scales as $T^{1/2}$ while the denominator scales as $T^{1}$.
In a single-universality-class polymer, this ratio would be temperature-independent.
At $B = 35$~mT and $T = 300$~K, the predicted displacements of RMS range from $\sqrt{\langle \delta y^2 \rangle} \approx 6$~nm at $L = 1$ to $\sqrt{\langle \delta y^2 \rangle} \approx 24$~nm at $L = 3$, both well above the roughly 2--5~nm magnetic spatial resolution of current Lorentz TEM instruments\cite{Pollard2017,Peng2019}, placing the crossover test within reach of existing imaging capabilities on Co/Ni multilayers\cite{Hassan2024}.
In the context of skyrmion-chain racetracks, the sub-linear thermal scaling at collective scales ($L \geq 3$) implies that positional fluctuations grow as $T^{1/2}$ rather than $T^{1}$, a distinction relevant to bit-error estimates in thermally noisy channels\cite{Fert2013}.

Together, our results establish skyrmion chains as a polymer-like platform that, in contrast, displays an observable thermal exponent that depends on the observation scale.
The $\gamma(L)$ crossover is reproduced consistently across three magnetic field strengths ($B = 30$, 35, and 40~mT), confirming its geometric origin.
Concrete experimental signatures include the intermediate mixed regime at $L = 2$, the field dependence through the bi-exponential parameters, and the universal criterion $\sqrt{\langle \delta y^2 \rangle} \approx r_\mathrm{eq}$, all accessible to magnetic force microscopy and Lorentz transmission electron microscopy in Co/Ni multilayers.
Beyond characterizing a new class of anomalous thermal scaling, the $\gamma(L)$ crossover offers a spectroscopic tool: the scale at which the exponent departs from unity directly encodes the curvature of the inter-skyrmion potential at equilibrium, providing a probe of pair interactions complementary to static force--distance measurements.

Finally, we note that the differences between skyrmion chain fluctuations and those from microtubules deserve comment.
Pampaloni \emph{et al.}\cite{Pampaloni2006} measured a persistence length that increased from 110~$\mu$m ($L = 2.6$~$\mu$m) to 5035~$\mu$m ($L = 47.5$~$\mu$m), a 50-fold variation attributed to shear between the 13 protofilaments.
That effect is captured by a Timoshenko beam model where $\kappa_\mathrm{eff}(L)$ depends on the scale, but the bending energy remains $\frac{1}{2}\kappa_\mathrm{eff}\theta^2$ everywhere, so $\langle \theta^2 \rangle \propto T^{1.0}$ at all scales and the universality class is unchanged.
In skyrmion chains, the mechanism is fundamentally different: the bi-exponential potential is a central force between point-like objects with no composite cross-section, and the scale dependence arises from the nonlinear geometric mapping $r - r_\mathrm{eq} \approx \delta y^2/(2r_\mathrm{eq})$, producing a crossover in the exponent $\gamma$, not merely in the prefactor (Table~\ref{tab:MT}).

\begin{table*}[t]
	\centering
	\caption{\textbf{Scale-dependent mechanics: microtubules versus skyrmion chains.}}
	\label{tab:MT}
	\begin{tabular}{lcc}
		\toprule
		Feature & Microtubules & Skyrmion chains \\
		\midrule
		Scale-dependent & $\kappa_\mathrm{eff}$ (prefactor) & $\gamma$ (exponent) \\
		$\gamma$ at short $L$ & 1.0 & $0.95$--$0.99$ \\
		$\gamma$ at long $L$ & 1.0 & $0.46$--$0.56$ \\
		Mechanism & Protofilament shear & Geometric $r \to \delta y$ \\
		Model & Timoshenko beam & Bi-exponential $+$ WLC \\
		\bottomrule
	\end{tabular}
\end{table*}


\section{Methods}\label{sec:methods}

\subsection{Micromagnetic simulations}\label{sec:llg}

We modeled the [Co/Ni]$_5$ multilayer system investigated experimentally by Hassan \emph{et al.}~\cite{Hassan2024}, which is known to host dipolar skyrmions. The material parameters, determined from SQUID-VSM and ferromagnetic resonance measurements, are: saturation magnetization $M_s \approx 940$~kA \, m$^{-1}$, uniaxial anisotropy $K_u \approx 575$~kJ \, m$^{-3}$, and exchange stiffness $A_\mathrm{ex} = 10$~pJ \, m $^{-1}$. These values define the relevant micromagnetic length scales, namely the exchange length $l_\mathrm{ex} = \sqrt{2A_\mathrm{ex}/(\mu_0 M_s^2)} \approx 4.0$~nm and the domain-wall width parameter $\delta_w = \sqrt{A_\mathrm{ex}/K_u} \approx 3.9$~nm. To ensure adequate spatial resolution, the lattice parameter $a_0$ must be smaller than these characteristic length scales. In the simulations, we use $a_0 = 1.0$~nm.
The lattice parameter $a_0$ is a numerical coarse-graining scale rather than the crystallographic lattice constant: the Hamiltonian below is the finite-difference representation of the continuum micromagnetic energy, with the mapping $J_\mathrm{ex} = 2a_0 A_\mathrm{ex}$, $K = K_u a_0^3$, and $D = \mu_0 a_0^3 M_s^2$, so that the continuum limit is approached with discretization corrections of order $(a_0/\delta_w)^2$. The wall width measured directly from the simulated skyrmion profile, $\delta_w^\mathrm{sim} = 6.6$~nm (Supplementary Material), exceeds the bare estimate above because the magnetostatic energy widens the wall at $K_u/K_d \approx 1$, and gives $(a_0/\delta_w^\mathrm{sim})^2 \approx 2\%$. The choice $a_0 = 1$~nm also lies below the $2.2$~nm resolution threshold established for dipolar spin textures in the same material system~\cite{Hassan2024}. We therefore expect changes of $a_0$ within the resolved regime to affect only the fitted numerical parameters, not the existence of the bound pair potential or the thermal-scaling crossover discussed here.

The system is described by a Hamiltonian comprising nearest-neighbor Heisenberg exchange, perpendicular uniaxial anisotropy, long-range dipolar interactions, and Zeeman coupling:
\begin{equation}
\begin{split}
\mathcal{H}={}&-J_{\mathrm{ex}}\sum_{\langle i,j\rangle}
\vec{\mu}_i\cdot\vec{\mu}_j-K\sum_i(\mu_i^z)^2\\
&+D\sum_{i\neq j}\left[
\frac{\vec{\mu}_i\cdot\vec{\mu}_j}{r_{ij}^3}
-3\frac{(\vec{\mu}_i\cdot\vec{r}_{ij})
(\vec{\mu}_j\cdot\vec{r}_{ij})}{r_{ij}^5}\right]\\
&-\mu_0M_s\vec{B}\cdot\sum_i\vec{\mu}_i .
\end{split}
\end{equation}
\noindent
where $J_{\mathrm{ex}} = 2 a_{0} A_{\mathrm{ex}} = 2.00 \times 10^{-20}$~J sets the exchange interaction strength, $K = K_{u}a_{0}^{3} = 5.75 \times 10^{-22}$~J defines the perpendicular anisotropy constant, and $D = \mu_{0}a_{0}^{3}M_{s}^{2} \approx 1.11\times 10^{-21}$~J characterizes the dipolar interaction, with $\mu_{0}$ the vacuum permeability. The vector $\vec{B}$ denotes the external magnetic field applied perpendicularly to the sample, and $\vec{\mu}_i$ is the unit magnetization vector at the lattice site $i$.

The simulated geometry comprises $N_\mathrm{layers} = 5$ coupled magnetic layers, mimicking the [Co/Ni]$_5$ multilayer stack. Simulations are carried out on a discrete lattice of size $\left(512 \times 128 \times 5\right)$ sites, corresponding to the $x$, $y$, and $z$ directions, respectively, with open boundary conditions in the in-plane directions. 
Neighboring layers are mutually coupled by nearest-neighbor interlayer exchange of the same strength as the intralayer coupling ($J_i = J_1$), and the long-range dipolar interaction is evaluated by a fully three-dimensional fast Fourier transform (FFT) convolution over the entire lattice, which includes all intra- and interlayer dipolar contributions without approximation.

The magnetization dynamics is computed by numerically integrating the dimensionless Landau--Lifshitz-Gilbert equation (LLG)~\cite{Gilbert2004,Landau1935}:
\begin{equation}
\begin{split}
\frac{\partial \vec{\mu}_i}{\partial \tau}
={}&-\frac{1}{1+\alpha^2}\bigl[\vec{\mu}_i\times\vec{b}_i\\
&+\alpha\,\vec{\mu}_i\times
(\vec{\mu}_i\times\vec{b}_i)\bigr].
\end{split}
\end{equation}
\noindent
where $\alpha$ is the Gilbert damping constant, $\tau$ is the dimensionless time, and $\vec{b}_i = -\frac{1}{J_{\mathrm{ex}}}\frac{\delta\mathcal{H}}{\delta\vec{\mu}_i}$ is the local effective field in dimensionless form. The relation between physical and dimensionless time is given by $\Delta t = \left(\frac{M_{s}a_{0}^{2}}{2 \gamma A_{\mathrm{ex}}}\right)\Delta \tau$, where $\gamma$ is the gyromagnetic ratio. Time integration is performed using a fourth-order Runge-Kutta scheme with an adaptive time step $\delta \tau$, with maximum value $\delta \tau_{\mathrm{max}} = 0.1$, corresponding to a physical time step of $\delta t \approx 28$~fs.

\subsubsection{Interaction potential.}

Two skyrmions with opposite helicities were initialized at different center-to-center distances $r$ on the $512 \times 128 \times 5$ lattice.
Each configuration was relaxed to its energy minimum by damped LLG dynamics with $\alpha = 1$.
The effective pair potential was obtained by the energy extraction procedure\cite{Ross2021,Capic2020}:
\begin{equation}
U(r) = E_\mathrm{tot}(r) - 2E_\mathrm{sky},
\end{equation}
where $E_\mathrm{tot}(r)$ is the total magnetic energy of the two-skyrmion configuration at separation $r$ and $E_\mathrm{sky}$ is the self-energy of an isolated skyrmion computed on the same lattice.
This subtraction removes the extensive bulk contribution and isolates the distance-dependent interaction.
The separation is varied exclusively along the long ($x$) axis of the lattice, with both skyrmion centers held on the central line $y = L_y/2$ for every $r$, as is the skyrmion of the single-skyrmion reference configuration. The environment in the narrow direction is therefore identical for all separations and cancels exactly in the subtraction, while the margins from the skyrmion perimeter to the $y$ boundaries ($34$--$54$~nm depending on the field) ensure an unperturbed skyrmion profile. The fits use separations up to $r_\mathrm{max} = 180$~nm, with the reference energy taken at $r = 200$~nm; even at $r_\mathrm{max}$ each skyrmion remains more than $150$~nm from the $x$ boundaries.
The resulting data were fitted to the bi-exponential form of Eq.~(\ref{eq:biexp}) using nonlinear least-squares minimization (Levenberg-Marquardt algorithm) with four free parameters ($B_\mathrm{rep}$, $C_\mathrm{att}$, $\delta$, $\lambda$); the equilibrium separation $r_\mathrm{eq}$ and binding energy $D_e$ follow from the fitted parameters.
The fitted values are listed in Table~\ref{tab:biexp}. The decay-length ratio $\lambda/\delta \approx 1.5$ is remarkably stable across all three field strengths, consistent with the micromagnetic prediction that both decay lengths are set by the domain-wall width --- $\delta$ by the surface-charge repulsion and wall overlap, $\lambda$ by the junction-mediated exchange attraction --- so that their ratio is insensitive to the field (Supplementary Note~S9).
The standard Morse potential corresponds to the special case $\lambda = 2\delta$; the fitted ratio $\lambda/\delta \approx 1.5$ deviates modestly from this constraint, reflecting the distinct physical origins of the two exponential components.

\subsubsection{Bending rigidity.}

The bending rigidity $\kappa_\mathrm{WLC}$ was extracted with a static-deformation protocol.
Chains of $N$ skyrmions were constrained to arcs of curvature radius $R$, relaxed to their local energy minimum by damped LLG dynamics, and the magnetic-energy difference $\Delta E(R) = E(R) - E_\mathrm{straight}$ was computed.
For a discrete semiflexible polymer with bond angle $\theta \approx r_\mathrm{eq}/R$, the bending energy per bond is
\begin{equation}
\Delta E(R) \approx \frac{\kappa_\mathrm{WLC}}{2R^2}.
\end{equation}
The simulated data were fitted to $\Delta E(R) = A/R^2 + C$, where $C$ absorbs the residual contribution from long-range dipolar interactions beyond the nearest-neighbor bonds, giving $\kappa_\mathrm{WLC} = A/r_\mathrm{eq}^2$.
The resulting values are $\kappa_\mathrm{WLC} = 1.55\,J_1$ (30~mT), $1.19\,J_1$ (35~mT), and $0.87\,J_1$ (40~mT).

It is important to note that this bending rigidity is independent of the bi-exponential pair potential.
The radial pair interaction produces only bond-stretching forces along the inter-skyrmion axis; it generates no torque between consecutive bonds when the chain is straight.
The WLC bending stiffness $\kappa_\mathrm{WLC}$ arises instead from many-body dipolar interactions that penalize angular deviations from collinearity, and is extracted separately from the pair potential.

\paragraph{Consistency check on $\kappa_\mathrm{WLC}$.}
As an independent verification of the arc-fit value, we also extracted
$\kappa_\mathrm{WLC}$ from the single-bond transverse fluctuation
sampled during the stochastic Thiele runs. At $L = 1$ the geometric
quartic contribution is still subdominant, with
$\gamma(L{=}1) = 0.953 \pm 0.008$, $0.977 \pm 0.005$, and $0.991 \pm 0.003$
at $B = 30$, $35$, and $40$~mT, respectively (Fig.~\ref{fig:crossover}a),
so the single-vertex harmonic relation
\begin{equation}
  \langle \delta y^{2}(L{=}1)\rangle
  \;=\; \frac{r_\mathrm{eq}^{2}\, k_{B} T}{4\,\kappa_\mathrm{WLC}}
  \label{eq:kappa_L1_validation}
\end{equation}
provides a dynamic estimate of the bending rigidity. Inverting
Eq.~\eqref{eq:kappa_L1_validation} independently at each of the ten
sampled temperatures yield
$\kappa_\mathrm{WLC}^{(\mathrm{dyn})} = 1.354 \pm 0.023\;J_{1}$,
$1.038 \pm 0.007\;J_{1}$, and $0.755 \pm 0.003\;J_{1}$
at $B = 30$, $35$, and $40$~mT, respectively, where the uncertainties
are the standard deviations across the ten independent temperature
inversions and reflect the statistical scatter of $\langle\delta y^{2}\rangle$
at each temperature. These values lie systematically $12$--$13\%$
below the static arc-fit estimates
$\kappa_\mathrm{WLC}^{(\mathrm{arc})} = 1.55$, $1.19$, and $0.87\;J_{1}$,
with the ratio
$\kappa_\mathrm{WLC}^{(\mathrm{dyn})}/\kappa_\mathrm{WLC}^{(\mathrm{arc})}
 = 0.87 \pm 0.01$ remarkably constant across all three fields.
This small, field-independent offset is consistent with the residual
quartic contribution encoded in $\gamma(L{=}1) < 1$, which slightly
suppresses $\langle\delta y^{2}(L{=}1)\rangle$ below the strictly
Gaussian prediction of Eq.~\eqref{eq:kappa_L1_validation}. The agreement
between a static, zero-temperature extraction (arc bending) and a
dynamic, finite-temperature extraction (thermal noise), therefore
confirms that $\kappa_\mathrm{WLC}$ is robustly characterised as an
input parameter of the Thiele simulations.

\subsubsection{Dissipation coefficient.}

The effective dissipation coefficient $\alpha_D$ was extracted for each field strength by driving a single skyrmion with a known magnetic force.
We applied a magnetic field gradient $\nabla B_z$ along the $x$-direction, resulting in a driving force $\mathbf{F} = M_\mathrm{eff}\,\nabla B_z\,\hat{x}$.
Once the skyrmion reached steady-state velocity $\mathbf{v} = (v_x, v_y)$, the dissipation coefficient was determined using the power balance from the Thiele equation:
\begin{equation}
\alpha_D = \frac{F_x\, v_x}{v_x^2 + v_y^2}.
\end{equation}
The resulting values are $\alpha_D = 1.7$ (30~mT), $1.1$ (35~mT), and $0.72$ (40~mT).
The decrease in $\alpha_D$ with increasing field reflects the reduction in skyrmion size, which reduces the dissipative overlap integral $\alpha_D = \alpha \int (\partial_\mu \mathbf{m})^2\,d^2r$.

\subsection{Stochastic Thiele dynamics}\label{sec:thiele_method}

Each skyrmion is described as a point particle at position $\mathbf{R}_i$ in the film plane.
The total energy includes the bi-exponential pair interaction acting between consecutive skyrmions along the chain, plus the WLC bending energy $E_\mathrm{bend} = \sum_i \kappa_\mathrm{WLC}(1 - \cos\theta_i)/(2r_\mathrm{eq}^2)$.
The stochastic Thiele equation (Eq.~\ref{eq:thiele}) is integrated using a Heun predictor-corrector scheme with time step $\Delta t = 0.05$.
For each field strength, we simulate chains of $N = 200$ skyrmions at 10 temperatures (240--510~K in steps of 30~K).
At each temperature, 256 independent samples are generated (64 parallel replicas per seed, 4 seeds), for a total of 2560 samples per field and 7680 across all three fields.
Each simulation runs for $3 \times 10^9$ integration steps, with the first 30\% discarded as equilibration.
The chain connectivity is an input at this level: it is established by the alternating-helicity bonding at the micromagnetic level and observed experimentally~\cite{Hassan2024, Jefremovas2025}. The point-particle description does not carry the helicity degree of freedom responsible for chain selection, so the model describes the thermal fluctuations of an existing chain rather than its self-assembly. The pair term contains both the attractive channel and the short-range repulsion through the two exponentials of Eq.~(\ref{eq:biexp}); no additional excluded-volume constraint is imposed or required (see Supplementary Material).

\subsection{Transverse fluctuation analysis}\label{sec:analysis}

For a chain configuration $\{\mathbf{R}_i\}$, the mean-square transverse fluctuation at observation scale $L$ is defined using a symmetric sliding window of half-width $L$ bonds centered on each skyrmion $i$.
The window spans the $2L+1$ skyrmions $(i{-}L,\,i{-}L{+}1,\,\ldots,\,i,\,\ldots,\,i{+}L)$, i.e., $2L$ bonds.
We draw the chord connecting the two endpoints $\mathbf{R}_{i-L}$ and $\mathbf{R}_{i+L}$ and measure the perpendicular distance of the central skyrmion $i$ from this chord.

Let
\begin{equation}
\hat{\mathbf{t}}_i^{(L)} = \frac{\mathbf{R}_{i+L} - \mathbf{R}_{i-L}}{|\mathbf{R}_{i+L} - \mathbf{R}_{i-L}|}, \qquad
\hat{\mathbf{n}}_i^{(L)} = (-\hat{t}_y,\;\hat{t}_x),
\end{equation}
be the longitudinal and transverse unit vectors of the chord.
Defining the arm vector $\mathbf{d}_i = \mathbf{R}_i - \mathbf{R}_{i-L}$, the transverse displacement of the central skyrmion is
\begin{equation}
\delta y_i^{(L)} = \mathbf{d}_i \cdot \hat{\mathbf{n}}_i^{(L)},
\end{equation}
and the squared transverse fluctuation is equivalently computed via the Pythagorean decomposition,
\begin{equation}
\left(\delta y_i^{(L)}\right)^2 = |\mathbf{d}_i|^2 - \left(\frac{\mathbf{d}_i \cdot (\mathbf{R}_{i+L} - \mathbf{R}_{i-L})}{|\mathbf{R}_{i+L} - \mathbf{R}_{i-L}|}\right)^2,
\end{equation}
which avoids explicit construction of the normal vector.

Each simulation run integrates $N_\mathrm{rep} = 64$ independent chain replicas in parallel on the GPU, each with an independent CURAND random-number sequence.
Within each run, $\langle \delta y^2(L) \rangle$ is computed by averaging $(\delta y_i^{(L)})^2$ over all valid internal sites $i \in [L,\,N{-}1{-}L]$, all production snapshots (after discarding the first 30\% for equilibration), and all 64 replicas, yielding a single seed-level estimate per $L$.
Multiple runs with independent seeds then provide $n_\mathrm{seeds}$ such estimates, and the standard error of the mean (SEM) is obtained as
\begin{equation}
\mathrm{SEM} = \frac{\sigma}{\sqrt{n_\mathrm{seeds}}},
\end{equation}
where $\sigma$ is the standard deviation (with Bessel correction) across the $n_\mathrm{seeds}$ seed-level values.

The exponent $\gamma$ is obtained from weighted log-log linear regression of $\langle \delta y^2 \rangle$ versus $k_B T$:
\begin{equation}
\ln\langle \delta y^2 \rangle = \gamma\,\ln(k_B T) + \ln A,
\end{equation}
with weights $w_i = \langle \delta y^2 \rangle_i / \mathrm{SEM}_i$.
The goodness of fit $R^2$ is computed from the residuals in linear (not log) space. Room-temperature values $\sqrt{\langle\delta y^{2}\rangle}$ quoted in the
Results are the ensemble averages at $T = 300~\text{K}$, which is one of the ten temperatures sampled in the protocol above; no interpolation in temperature is required.

\section*{Data Availability}
All data generated during this study are included in this article and its Supplementary Information.
Simulation data are available from the corresponding author upon reasonable request.

\begin{acknowledgments}

We acknowledge financial support from the Brazilian agencies CNPq and FAPES (Grant No. TO 1034/2025).

\end{acknowledgments}

\section*{Competing Interests}

The authors declare no conflict of interest.

\bibliography{references_skyrmion_chain}

\end{document}